\documentclass[a4paper]{jpconf}
\usepackage{graphicx}
\usepackage{lineno}

\newcommand{\ud}{\mathrm{d}}

\begin{document}
\title{The charm of hot matter - charmonium and open charm measurements in Pb--Pb collisions with ALICE at the LHC}

\author{Anton Andronic, for the ALICE Collaboration}

\address{EMMI \& GSI Research Division, Planckstr. 1, 64291 Darmstadt, Germany}

\ead{A.Andronic@gsi.de}

\begin{abstract}
We present selected results obtained by the ALICE Collaboration concerning 
charmed hadron and quarkonium production in Pb--Pb collisions at the LHC.
\end{abstract}

\section{Introduction}

The goal of high-energy nucleus-nucleus collisions is to produce and 
characterize a state of nuclear (QCD) matter at (energy) densities well above 
the nuclear ground state ($\varepsilon_0\simeq$0.15 GeV/fm$^3$).
At high densities and/or at high temperatures one expects 
\cite{Collins:1974ky,Cabibbo:1975ig}
that quarks are no longer confined in protons and neutrons but move
freely over distances larger than the size of the nucleon 
($\simeq$1 fm=10$^{-15}$ m).
Employing Quantum Chromo-Dynamics calculations on a space-time lattice, 
a deconfinement phase transition for an energy density of about 1 GeV/fm$^3$ 
was predicted (see \cite{Karsch:2001cy} for an early review).
The characterization of quark-gluon matter in terms of its equation of state 
(EoS, relating pressure to energy) and of its transport properties (such as
viscosity) and delineating its phase diagram is a major ongoing research effort
in collisions of nuclei at high energies.

Heavy quarks (charm and bottom) are a prominent part of the so-called
 ``hard probes'', observables characterized by an energy (mass) scale much 
larger than the temperature of the deconfined medium (of a few hundred MeV).
These observables are produced in primary hard scattering processes and their 
yields are calculable in elementary hadronic collisions with perturbative QCD 
techniques \cite{ALICE:2011aa,Kniehl:2012ti,Cacciari:2012ny,Maciula:2013wg}.
The production of hard probes in nucleus-nucleus (AA) collisions is quantified 
by the nuclear modification factor:
$R_{\mathrm{AA}}=\frac{\ud^2 N_{\mathrm{AA}}/\ud p_{\mathrm T}\ud y}{\langle N_{\mathrm{coll}}\rangle\cdot\ud^2 N_{\mathrm{pp}}/\ud p_{\mathrm T}\ud y},$
where $\ud^2 N/\ud p_{\mathrm T}\ud y$ denotes the transverse momentum 
($p_{\mathrm T}$) and rapidity ($y$) differential yield of a given 
observable measured in AA or pp collisions 
and $\langle N_{\mathrm{coll}}\rangle$ is the average number of nucleon-nucleon collisions over 
the given centrality interval of AA collisions;
$\langle N_{\mathrm{coll}}\rangle$ is calculated using the Glauber model \cite{Miller:2007ri}.
Also calculated within this geometric model is the number of participating
nucleons, $\langle N_{\mathrm{part}}\rangle$, used in the following for study 
of centrality dependence of $R_{\mathrm{AA}}$.

The $p_{\mathrm T}$ spectra of hadrons carrying light 
quarks and their azimuthal distributions with respect to the reaction plane
(defined by the beam axis and the impact parameter of the colliding nuclei)
exhibit features of collective flow and are described well by hydrodynamical 
models (see a recent review in \cite{Heinz:2013th}).
In non-central collisions, the initial approximately elliptic shape 
(in the transverse plane) of the overlap zone of the colliding nuclei leads, 
through the initial gradients of the energy density (or pressure), to 
anisotropic spatial (azimuthal angular) emission of hadrons.
This is quantified by the second order (quadrupole) Fourier coefficient
$v_2=\langle \cos(2\phi)\rangle$, where $\phi$ is the azimuthal angle with 
respect to the reaction plane.

The study of (deconfined) QCD matter has entered a new era in year 2010 with 
the advent of Pb--Pb collisions at the Large Hadron Collider (LHC), 
delivering the largest ever collision energy, $\sqrt{s_{\mathrm{NN}}}$=2.76 TeV,
more than a factor of 10 larger than previously available. We present selected 
results obtained by the ALICE Collaboration concerning charmed hadron and 
quarkonium production.
The ALICE detector is described in \cite{Aamodt:2008zz}.  
The results presented below correspond to minimum-bias Pb-Pb run of the year 
2010 and to the high-statistics Pb-Pb run of the year 2011. An integrated 
luminosity $L_{int}\simeq$30 $\mu$b$^{-1}$ was collected in 2011 with centrality 
triggers for the ALICE Central Barrel at midrapidity ($|y|<0.9$),
while $L_{int}\simeq$70 $\mu$b$^{-1}$ was collected with (di)muon triggers in 
the forward Muon Spectrometer ($2.5<y<4$).

\section{Open charm hadrons} 

Produced very early in the collision, at times $t\simeq 1/2m_q$, heavy quarks 
experience the full evolution of the hot fireball of partons. 
At variance with light quarks, which can be thermally produced, heavy quarks 
maintain their identity through the hot stage of the collision.
They help to address the question if parton energy loss (by gluon radiation) 
exhibits the expected quark mass pattern. 
The theoretical expectation is that heavy quarks (charm and bottom) lose 
less energy (by gluon radiation) compared  to lighter (up, down, strange) 
quarks \cite{Dokshitzer:2001zm}. 
The related question, often asked, is whether heavy quarks thermalize 
alongside the light quarks and gluons.
The measurements at the LHC were eagerly awaited in order to establish in a new
energy regime the observation at RHIC, at a center of mass energy per nucleon 
pair $\sqrt{s_{\mathrm{NN}}}$ = 200 GeV, of heavy quark energy loss and collective 
flow \cite{Adare:2010de}.

\begin{figure}[htb]
\begin{tabular}{ccc}
\begin{minipage}{0.33\textwidth}
\hspace{-0.5cm}\includegraphics[width=1.1\textwidth]{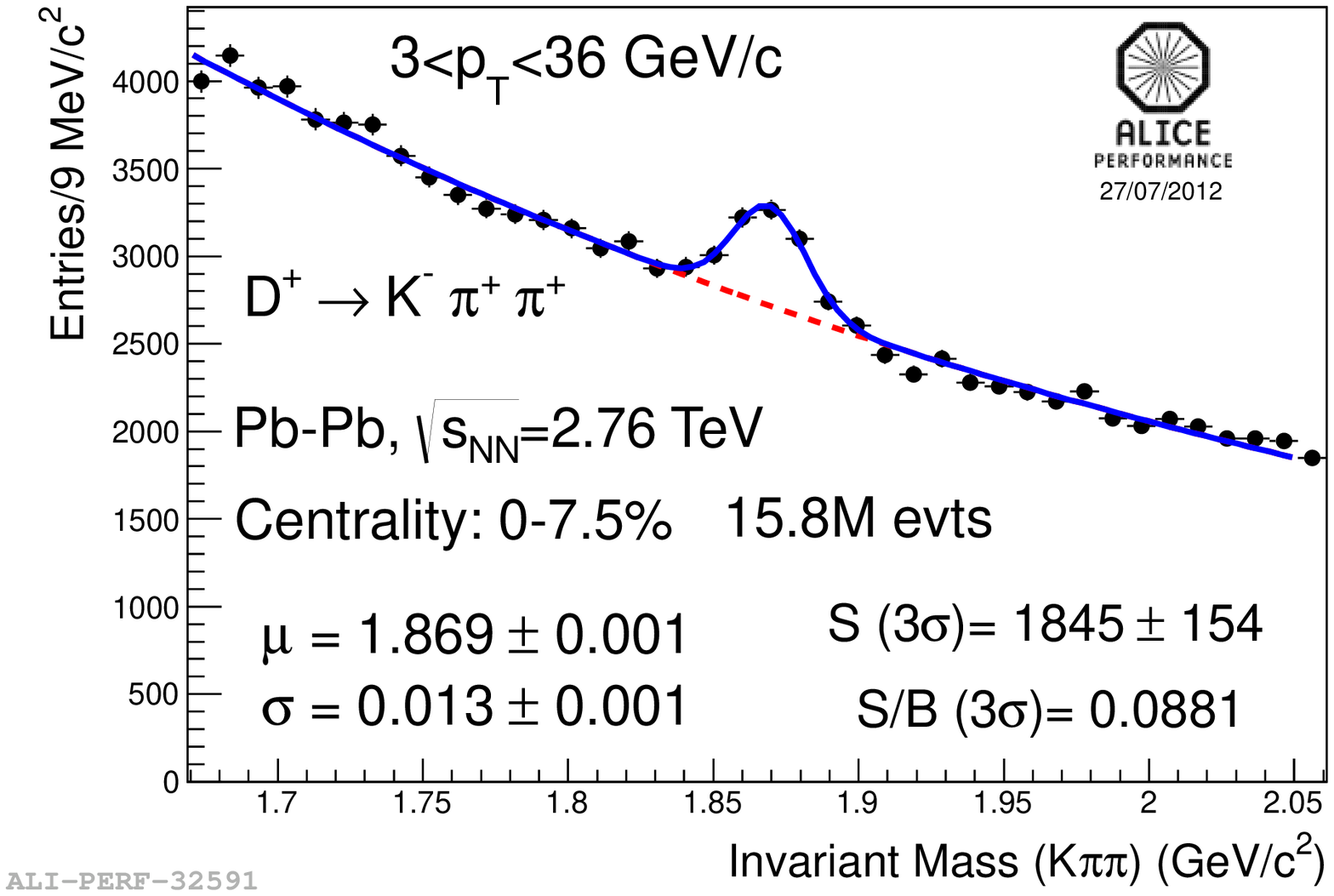}
\end{minipage} & \begin{minipage}{.33\textwidth}
\hspace{-0.5cm}\includegraphics[width=1.1\textwidth]{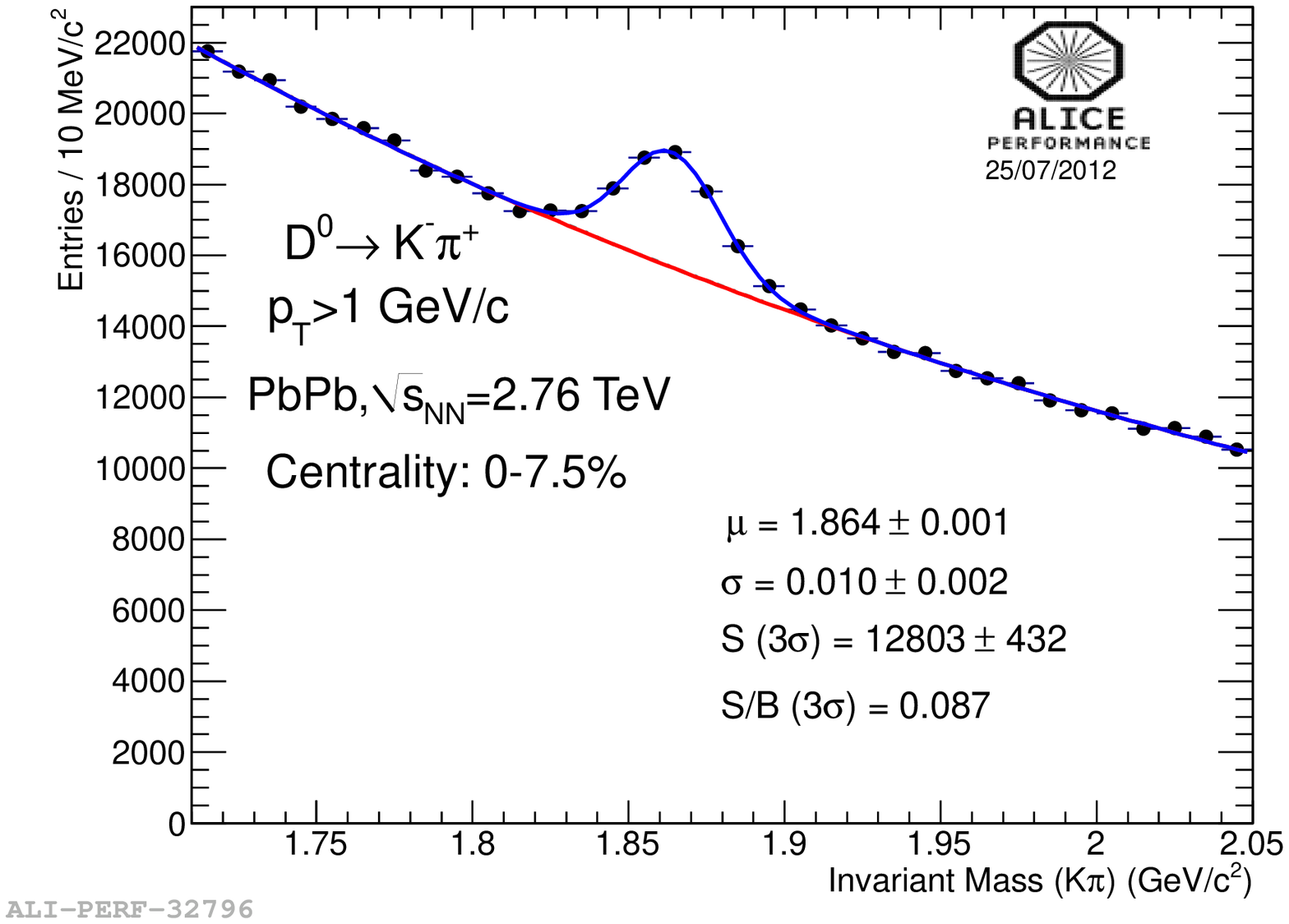}
\end{minipage} & \begin{minipage}{.33\textwidth}
\hspace{-0.5cm}{\includegraphics[width=1.\textwidth]{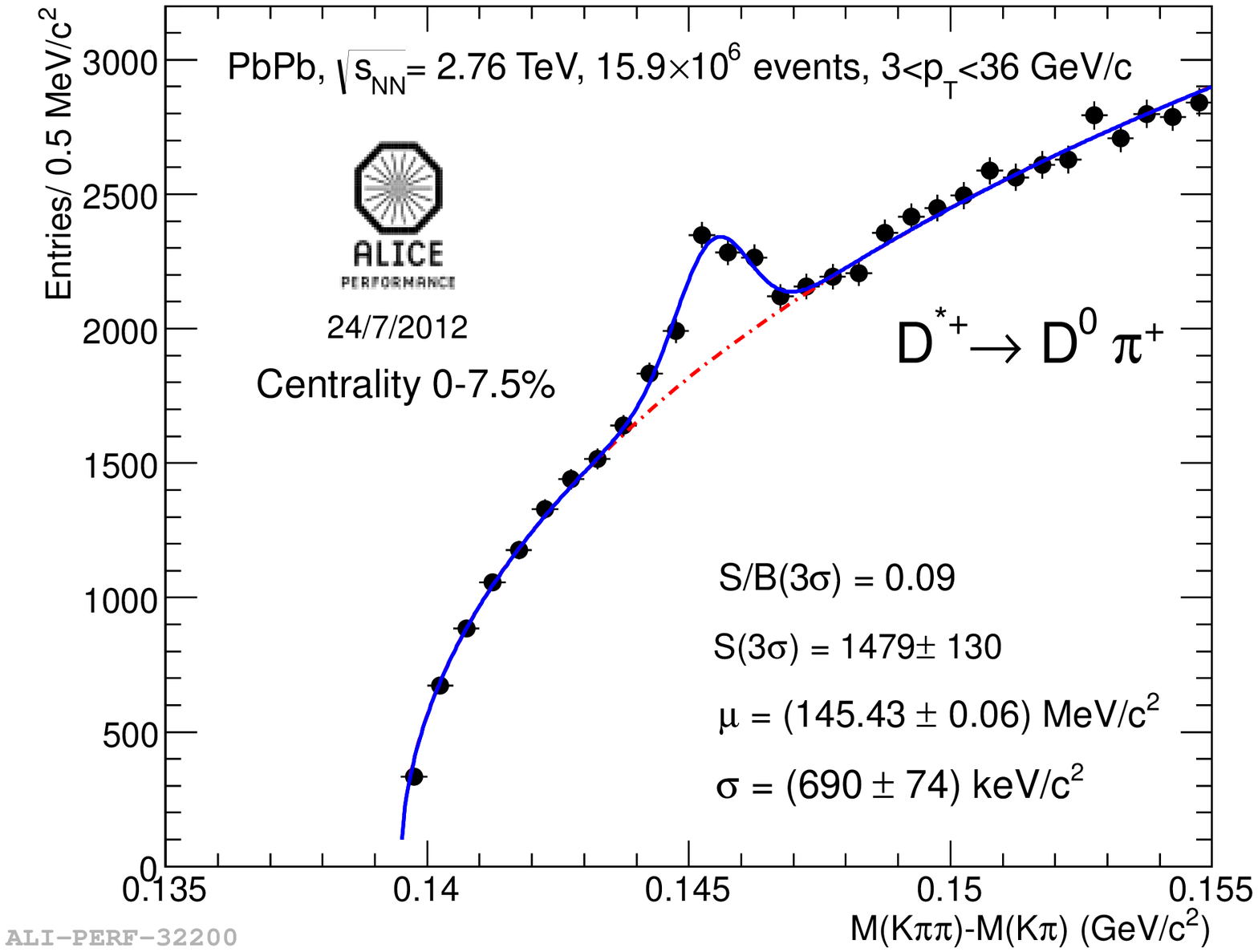}}
\end{minipage} \end{tabular}
\caption{\label{D1}
Invariant mass distributions for charmed meson candidates for central 
(0-7.5\% centrality) Pb-Pb collisions. Particles and antiparticles are included.}
\end{figure}

The production of hadrons carrying open charm is studied in ALICE via the 
direct reconstruction of hadronic decays \cite{ALICE:2012ab} at midrapidity, 
$|y|<0.5$, and via the measurement of decay electrons \cite{delValle:2012qw} 
(at midrapidity) or muons \cite{Abelev:2012qh} (at forward rapidity, $2.5<y<4$).
Invariant mass distributions for charmed meson candidates are shown in 
Fig.~\ref{D1} for central (0-7.5\% centrality) Pb-Pb collisions.
Owing to the very good resolution of track impact parameter in the transverse 
plane (65 $\mu$m for $p_{\mathrm T}=1$ GeV/$c$ and reaching asymptotically 
20 $\mu$m) and to the good hadron identification performance, 
a very good signal to background ratio S/B of about 10\% is 
reached for various D-meson species. The very good momentum resolution
of ALICE at low $p_{\mathrm T}$ leads to invariant mass resolutions
around 10 MeV/$c^2$ for the D$^+$ and D$^0$ mesons.

\begin{figure}[htb]
\begin{tabular}{cc}
\begin{minipage}{0.47\textwidth}
\includegraphics[width=.98\textwidth]{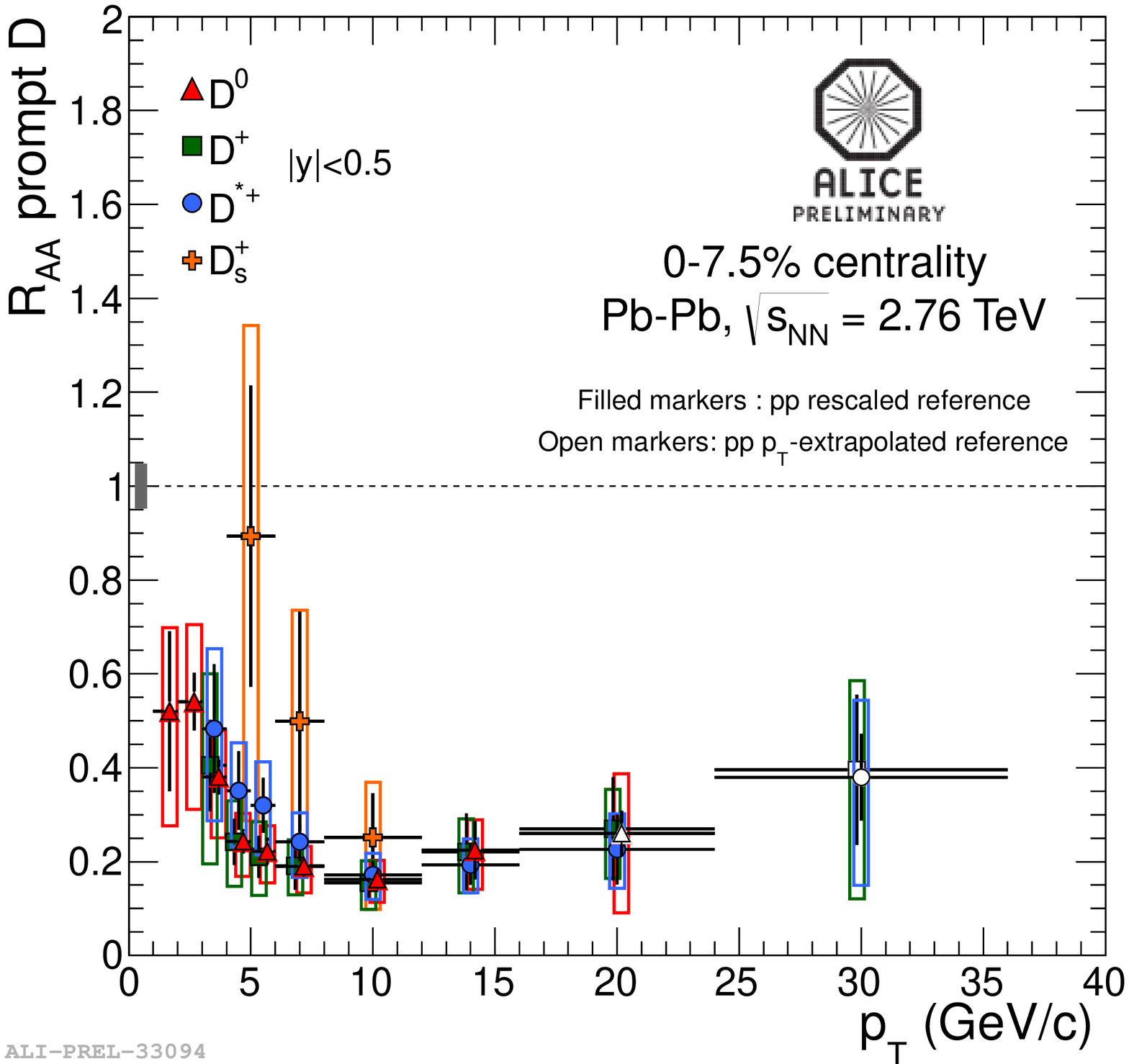}
\end{minipage} & \begin{minipage}{.47\textwidth}
\includegraphics[width=1.17\textwidth]{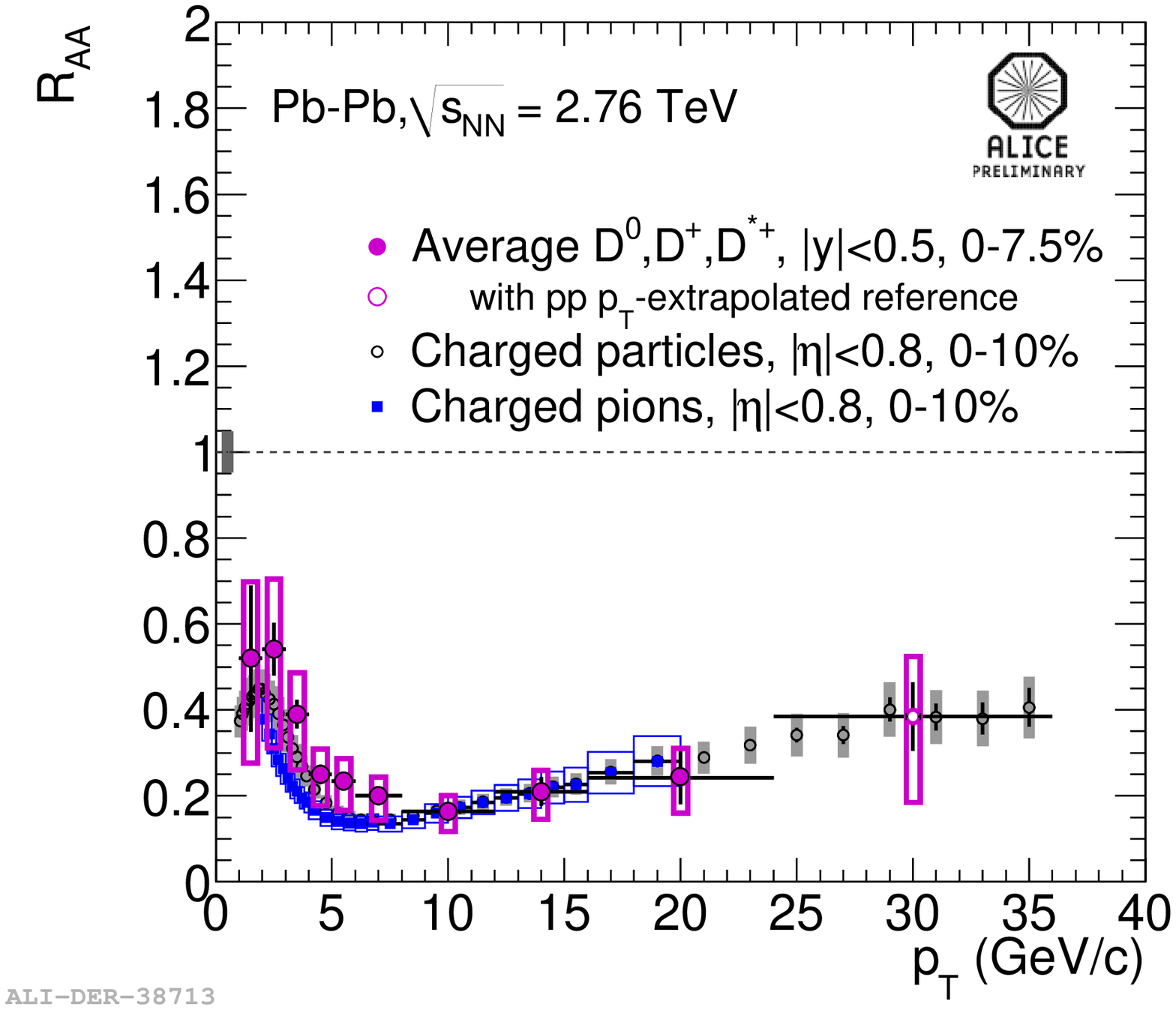}
\end{minipage}  \end{tabular}
\caption{\label{D2}
The transverse momentum dependence of the nuclear modification factor of 
prompt D mesons (particles and antiparticles) in central (0-7.5\% centrality) 
Pb--Pb collisions (left panel). 
The right panel shows the average values of $R_{\mathrm{AA}}$ for D$^+$, D$^0$, 
and  D$^{*+}$ meson species in comparison to data for all charged particles and 
for charged pions (for 0-10\% centrality).}
\end{figure}

The transverse momentum dependence of the nuclear modification factor 
$R_{\mathrm{AA}}$ of D mesons in central Pb-Pb collisions is shown in 
Fig.~\ref{D2} \cite{delValle:2012qw}.
The left panel shows the measurements for three D-meson species, D$^+$, D$^0$, 
and D$^{*+}$, while the right panel shows the average values of these species in 
comparison to data for all charged particles and for charged pions (which are 
for 0-10\% centrality). The similar $R_{\mathrm{AA}}$ values of D mesons and of 
pions (or charged particles), well below unity, indicate that the energy loss 
of charm quarks is of similar magnitude compared to that of lighter quarks 
or gluons.
The first measurement of $R_{\mathrm{AA}}$ for D$_s$ mesons in heavy-ion 
collisions \cite{Innocenti:2012ds} is also shown in Fig.~\ref{D2} (left panel).
In the highest measured $p_{\mathrm T}$ bin (8-12 GeV/$c$), the $R_{\mathrm{AA}}$ 
for D$_s$ mesons is compatible with that of non-strange charmed mesons. 
At lower $p_{\mathrm T}$, the $R_{\mathrm{AA}}$ of D$_s$ mesons seems to increase, 
but with the current statistical and systematic uncertainties no conclusion 
can be drawn on the expected enhancement of D$_s$mesons with respect to 
non-strange D mesons \cite{He:2012df}.

\begin{figure}[htb]
\centerline{\includegraphics[width=.6\textwidth]{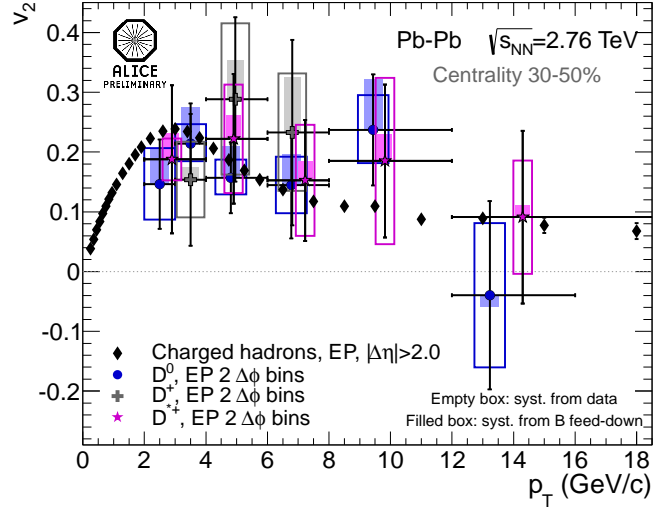}}
\caption{\label{D3}
The transverse momentum dependence of the elliptic flow coefficient $v_2$ of 
charmed mesons for the centrality interval 30-50\%. The data for inclusive 
charged hadrons is included for comparison.}
\end{figure}

The transverse momentum dependence of the elliptic flow coefficient $v_2$ of 
charmed mesons \cite{delValle:2012qw} is shown in Fig.~\ref{D3} together with 
that of inclusive charged hadrons for the centrality interval 30-50\%. 
Despite the large uncertainties, the challenging measurement of $v_2$ for 
D mesons shows a significant anisotropy, suggesting that charm
quarks may take part in the collective motion built up at the 
quark level in the deconfined stage.

\begin{figure}[htb]
\begin{tabular}{cc} 
\begin{minipage}{.5\textwidth}
\includegraphics[width=1.\textwidth]{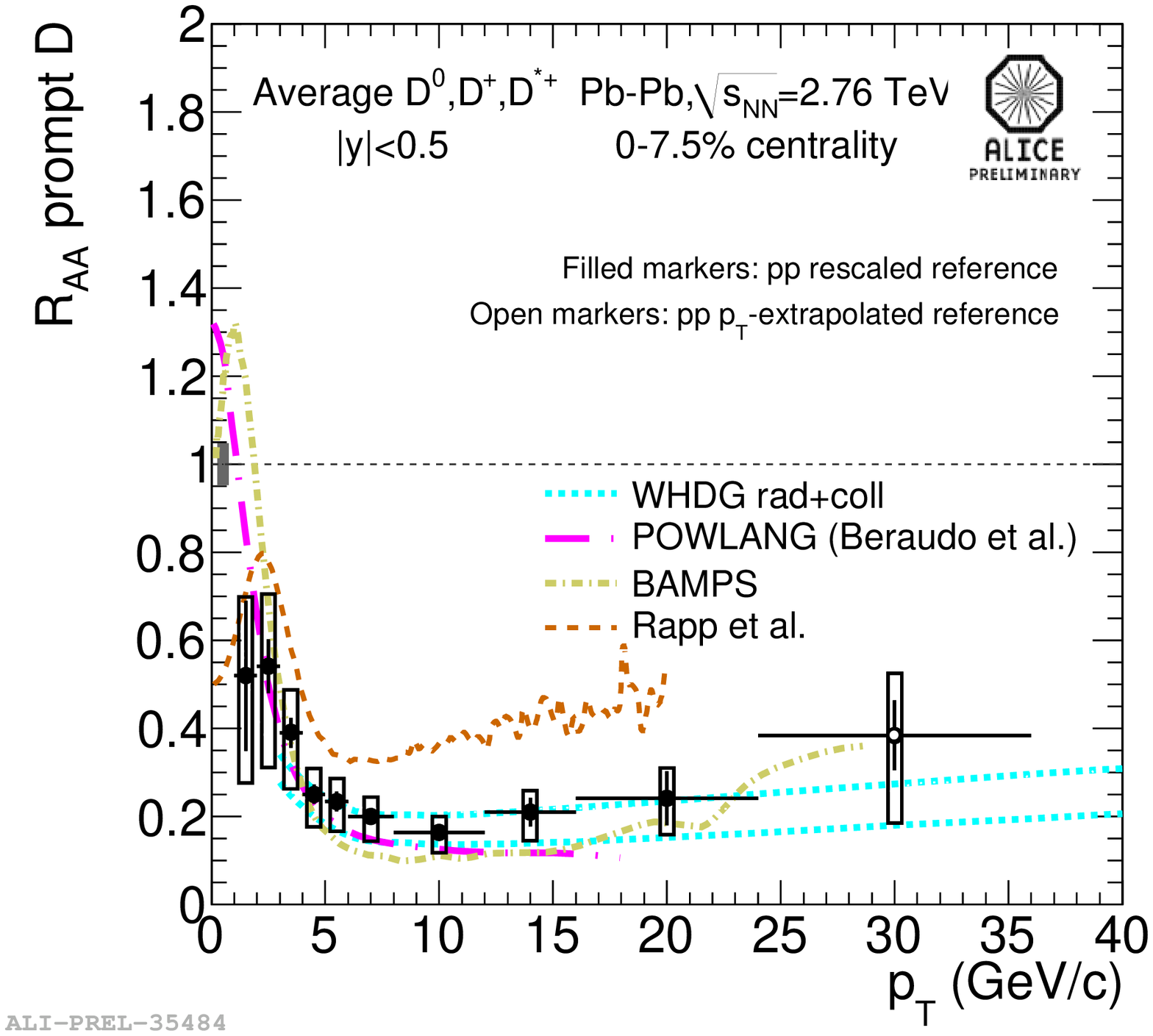}
\end{minipage} &\begin{minipage}{.5\textwidth}
\includegraphics[width=1.\textwidth,height=.9\textwidth]{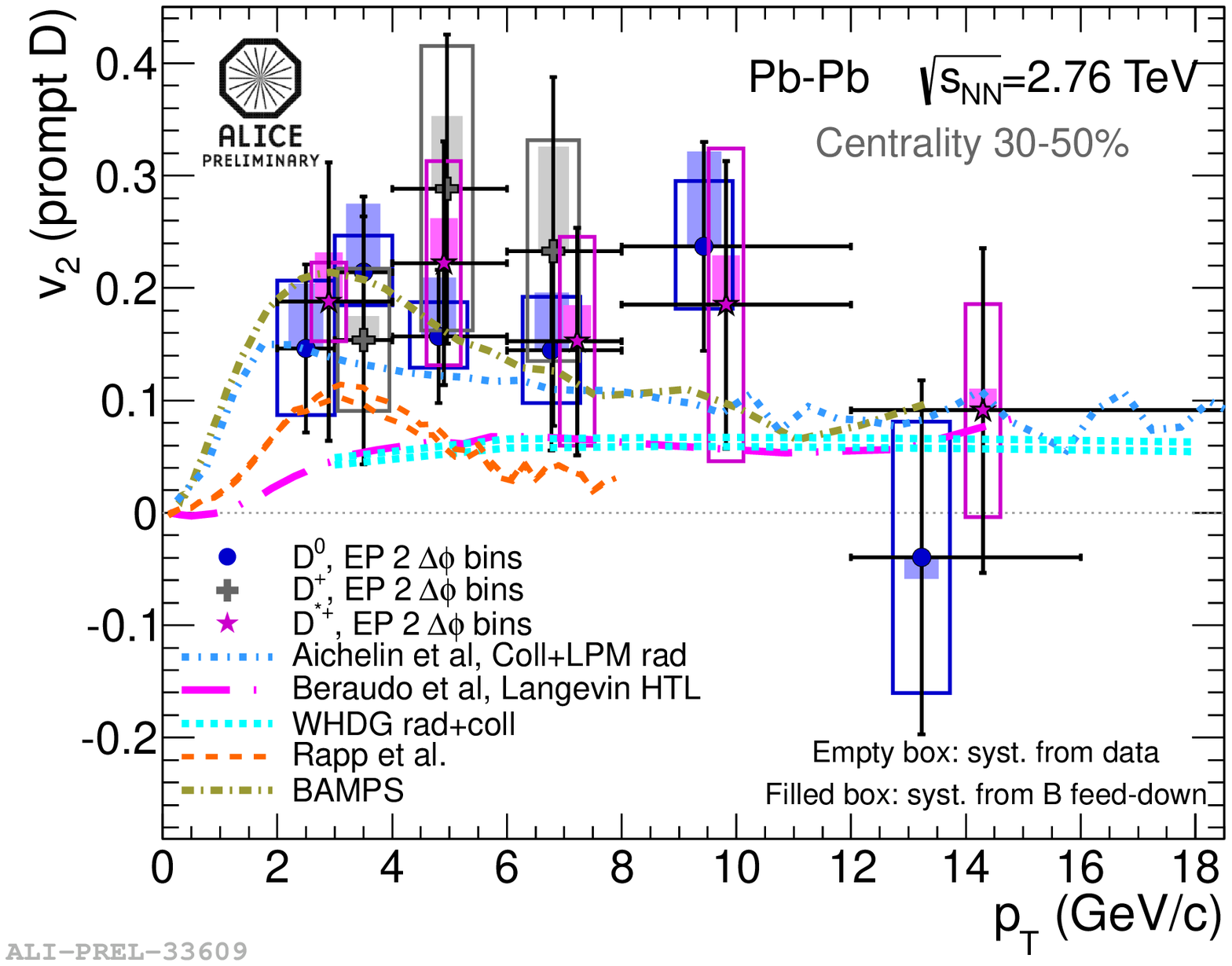}
\end{minipage} \end{tabular}
\caption{\label{D4} 
Transverse momentum dependence of $R_{\mathrm{AA}}$ (left panel, central collisions) 
and $v_2$ (right panel, 30-50\% centrality) of D mesons. The mesurements are 
compared to model predictions (see text).}
\end{figure}

The $R_{\mathrm{AA}}$ and $v_2$ results are confronted with model predictions in 
Fig.~\ref{D4}.
The theoretical models \cite{Gossiaux:2009mk,Greco:2011fs,Sharma:2009hn,Alberico:2011zy,Buzzatti:2011vt,Uphoff:2012it,Horowitz:2012cf}
implement parton energy loss via elastic (collisional) and radiative mechanisms.
The models are able to describe the features of the data, but a 
quantitative description of both energy loss and elliptic flow remains 
challenging for all models. 
As noted in \cite{Horowitz:2012cf}, an increase of $R_{\mathrm{AA}}$ as a function of
$p_{\mathrm T}$ is generic for a constant fractional parton energy loss and for
a parton spectrum of power-law form, with the power exponent increasing as a function
of $p_{\mathrm T}$.

\begin{figure}[hbt]
\begin{tabular}{cc} 
\begin{minipage}{.5\textwidth}
\includegraphics[width=.94\textwidth]{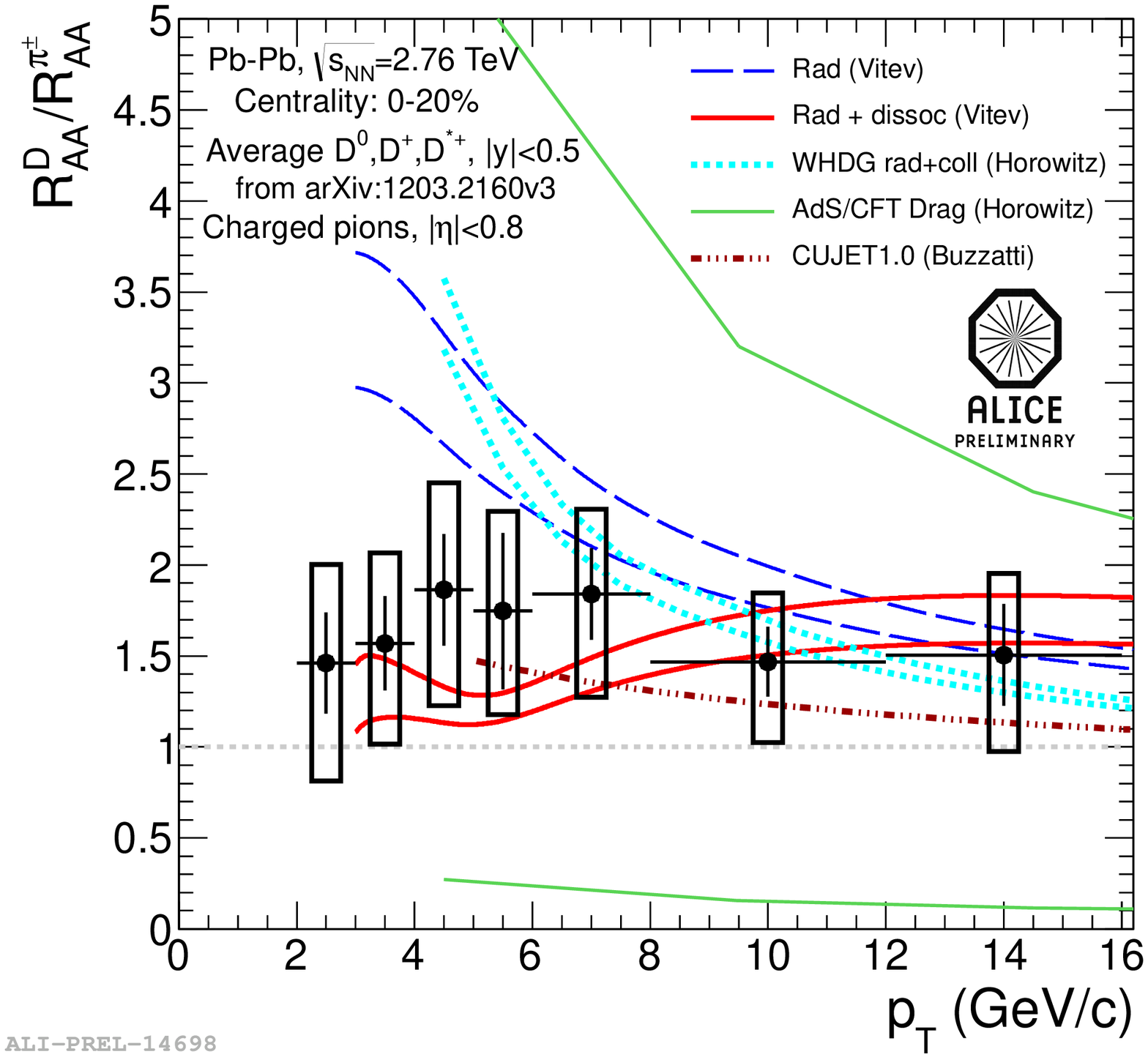}
\end{minipage} &\begin{minipage}{.5\textwidth}
\includegraphics[width=.94\textwidth,height=.85\textwidth]{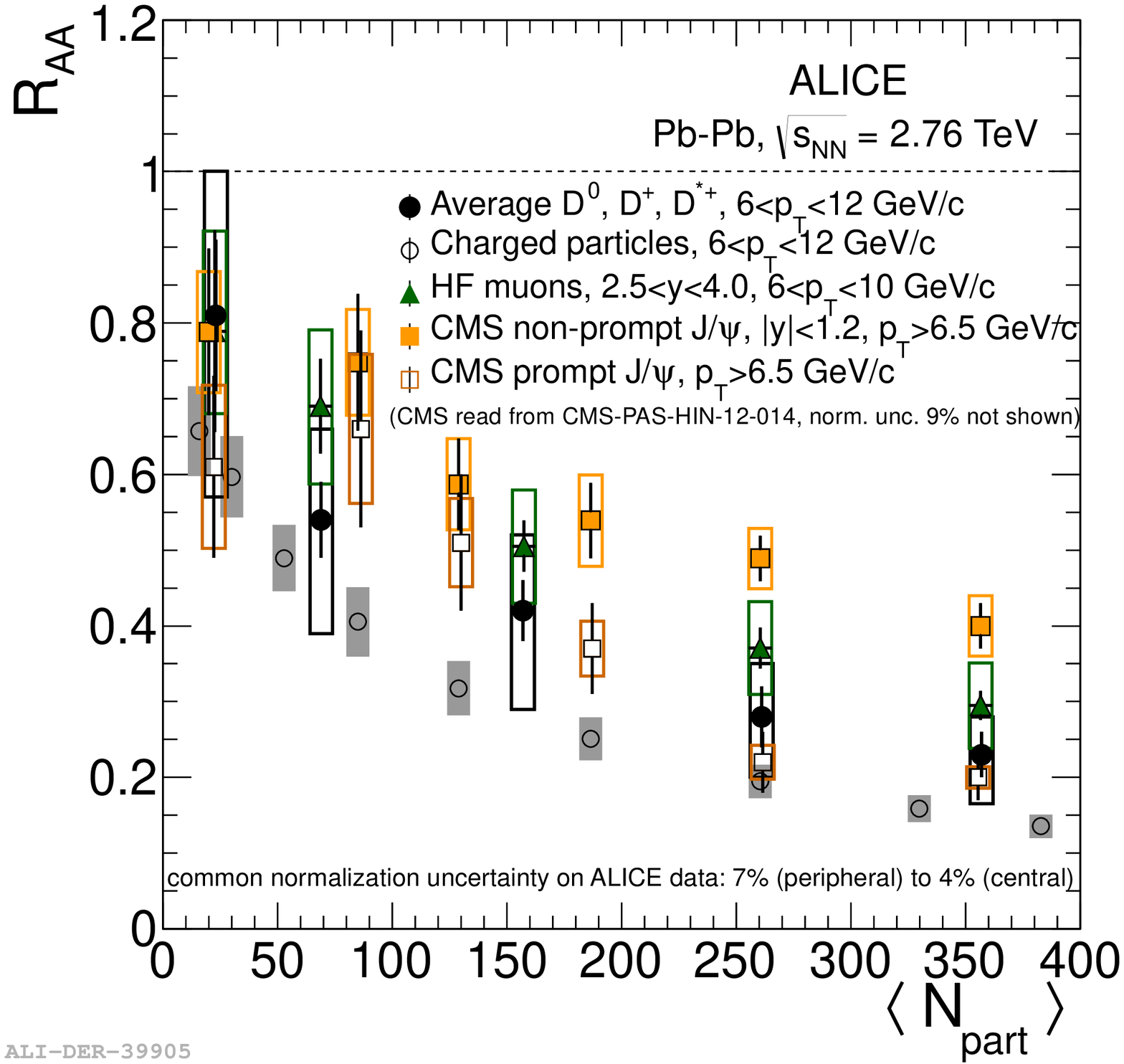}
\end{minipage} \end{tabular}
\caption{\label{D5}
Left panel: transverse momentum dependence of the ratio of $R_{\mathrm{AA}}$ of 
D mesons and pions, data and model predictions. 
Right panel: centrality dependence of  $R_{\mathrm{AA}}$ for hadrons carrying 
light and heavy quarks (see text).
}
\end{figure}

A further challenge for theory is the consistent description of energy loss for 
light \cite{Abelev:2012eq} and heavy quarks \cite{ALICE:2012ab}, illustrated in 
Fig.~\ref{D5} (left panel). 
Partonic energy loss models achieve a good description at high $p_{\mathrm T}$ 
while the low $p_{\mathrm T}$ region is not described well.
The data suggest $R_{\mathrm{AA}}^{D}>R_{\mathrm{AA}}^{\pi}$, as expected. Further 
hints on the quark flavor dependence of the nuclear modification factor are
given in  Fig.~\ref{D5} (right panel) for intermediate $p_{\mathrm T}$ values 
(where the $R_{\mathrm{AA}}$ values exhibit a shallow minimum, see Fig.~\ref{D4}). 
The ALICE data on light flavors \cite{Abelev:2012eq} and charmed hadrons 
\cite{delValle:2012qw} are compared to the measurements of primordial J/$\psi$ 
and J/$\psi$ from decays of B hadrons \cite{CMS-PAS-HIN-12-014}.
Although the details of such a comparison need to be carefully considered
\cite{Horowitz:2012cf}, the data suggest the expected hierarchy of flavor dependence
of in-medium parton energy loss.
A further challenge to theory is the extraction of the heavy quark 
(momentum) diffusion coefficient, a quantity which was recently calculated 
within lattice QCD \cite{Banerjee:2011ra}.

\section{Quarkonium} 

Among the various suggested probes of deconfinement, charmonium plays a 
distinctive role. It is the first hadron for which a clear mechanism of 
suppression in the QGP was proposed early on, based on the color analogue of 
Debye screening \cite{Matsui:1986dk} (see \cite{Kluberg:2009wc} for a 
recent review).
A suppression compared to pp collisions was observed in pA collisions
\cite{Arnaldi:2010ky}, 
which was understood as a destruction of the pre-resonant $c\bar{c}$ state 
by the nucleons of the colliding nuclei. 
The measurements in Pb-Pb at the SPS ($\sqrt{s_{\mathrm{NN}}}$ = 17.3 GeV)
\cite{Alessandro:2004ap}
and in Au-Au at RHIC \cite{Adare:2006ns}
demonstrated an "anomalous" suppression, attributed to the dense and hot QCD 
matter. The theoretical interpretation was a sequential suppression 
\cite{Digal:2001ue,Karsch:2005nk} of various charmonium states as a function 
of energy density (temperature) of the QGP.
In this picture, the J/$\psi$ meson survives in the QGP and only the 
contribution from the $\psi(2S)$ and $\chi_c$ charmonium states (leading to 
about 10\% and 30\% of the total J/$\psi$ yield, respectively) vanishes as 
a consequence of the melting of $\psi(2S)$ and $\chi_{c1,2}$ states.
As appealing as this interpretation is, it was questioned by
the non-zero $\psi(2S)$ yield measured at the SPS 
\cite{Alessandro:2006ju} and also later by the re-evaluation of cold 
nuclear matter suppression derived from p-A measurements \cite{Arnaldi:2010ky}.

The $\psi(2S)$ yield relative to the yield of J/$\psi$ was noted 
earlier \cite{Sorge:1997bg,BraunMunzinger:2000px} to correspond, for central 
collisions, to the chemical freeze-out temperature derived from fits to 
other hadron abundances.
This led to the idea of statistical hadronization of charm quarks in 
nucleus-nucleus collisions \cite{BraunMunzinger:2000px}. 
In this model, the charm quarks produced in initial hard collisions 
thermalize in the QGP and are ``distributed'' into hadrons at chemical 
freeze-out. 
All charmonium states are assumed to be fully melted (or, more precisely, 
not formed at all) in the QGP and produced, together with all other hadrons, 
exclusively at chemical freeze-out.
This model \cite{Andronic:2006ky,Andronic:2007bi} (see also a recent 
review \cite{BraunMunzinger:2009ih}) has gained support from experimental 
data at RHIC \cite{Adare:2006ns,Adare:2011yf}.
The model predicted a notably different pattern of J/$\psi$ production 
at the LHC.
Depending on the charm production cross section, even an enhanced
production relative to pp collisions could be expected at the LHC in
central Pb--Pb collisions
\cite{Andronic:2007bi} (see \cite{Andronic:2011yq} for further predictions).
The statistical model predicted \cite{Andronic:2009sv} an energy-independent 
relative production ratio $\psi(2S)$/J/$\psi$ about 4 times smaller in
(central) AA collisions compared to pp.

Proposed at the same time as the statistical hadronization model, the 
idea of kinetic recombination of charm and anti-charm quarks in the QGP 
\cite{Thews:2000rj} is an alternative quarkonium production mechanism.
In this model (see \cite{Liu:2009nb,Zhao:2011cv} for recent results), 
a continuous dissociation and regeneration of charmonium takes place in the 
QGP over its entire lifetime. Besides the charm production cross section, 
this model has as input parameters the time dependence of the 
temperature of the fireball as well as relevant cross sections and assumptions 
on the melting scenarios of charmonium states.
Important observables, like the transverse momentum dependence 
of production yields and of elliptic flow can be calculated within 
the transport (kinetic) models.
It predicted a rather small J/$\psi$ regeneration component at RHIC
and a sizable one at the LHC.

The measurement of J/$\psi$ production in Pb-Pb collisions at the LHC was 
expected to provide a definitive answer on the question of (re)generation.
The first data, measured at high-$p_{\mathrm T}$ by ATLAS \cite{Aad:2010aa}
and CMS \cite{Chatrchyan:2012np},
showed a pronounced suppression with respect to pp collisions.
Subsequently, it was seen that this J/$\psi$ 
suppression is of the same magnitude as that of open-charm hadrons 
\cite{ALICE:2012ab}, as shown above (Fig.~\ref{D4}, right panel).
This may indicate that the high-$p_{\mathrm T}$ charm quarks that form either 
D or J/$\psi$ mesons had the same dynamics including a thermalization 
stage and a late hadronization.

\begin{figure}[htb]
\begin{tabular}{cc} 
\begin{minipage}{.5\textwidth}
\includegraphics[width=1.\textwidth]{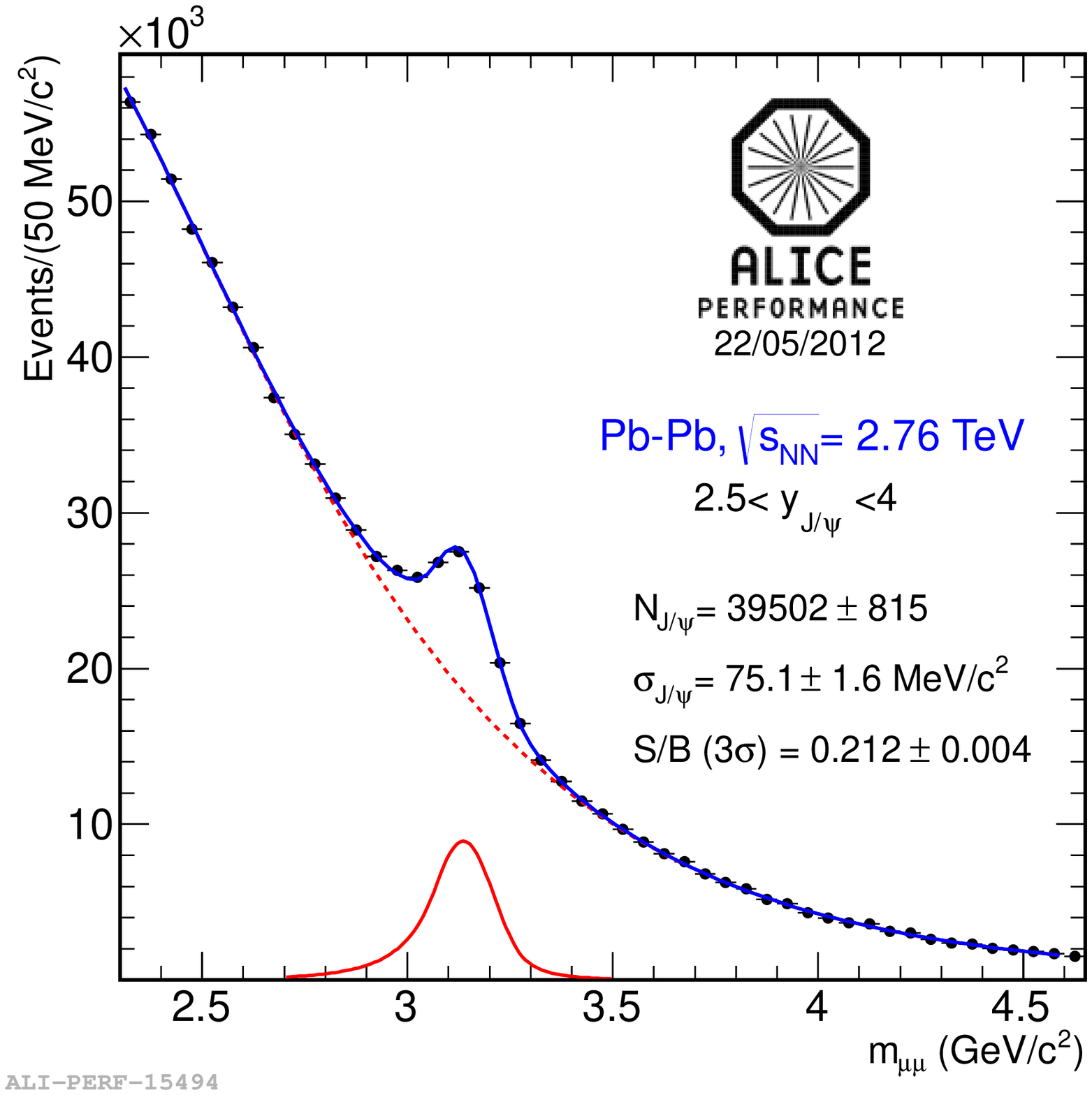}
\end{minipage} & \begin{minipage}{.5\textwidth}
\includegraphics[width=.87\textwidth]{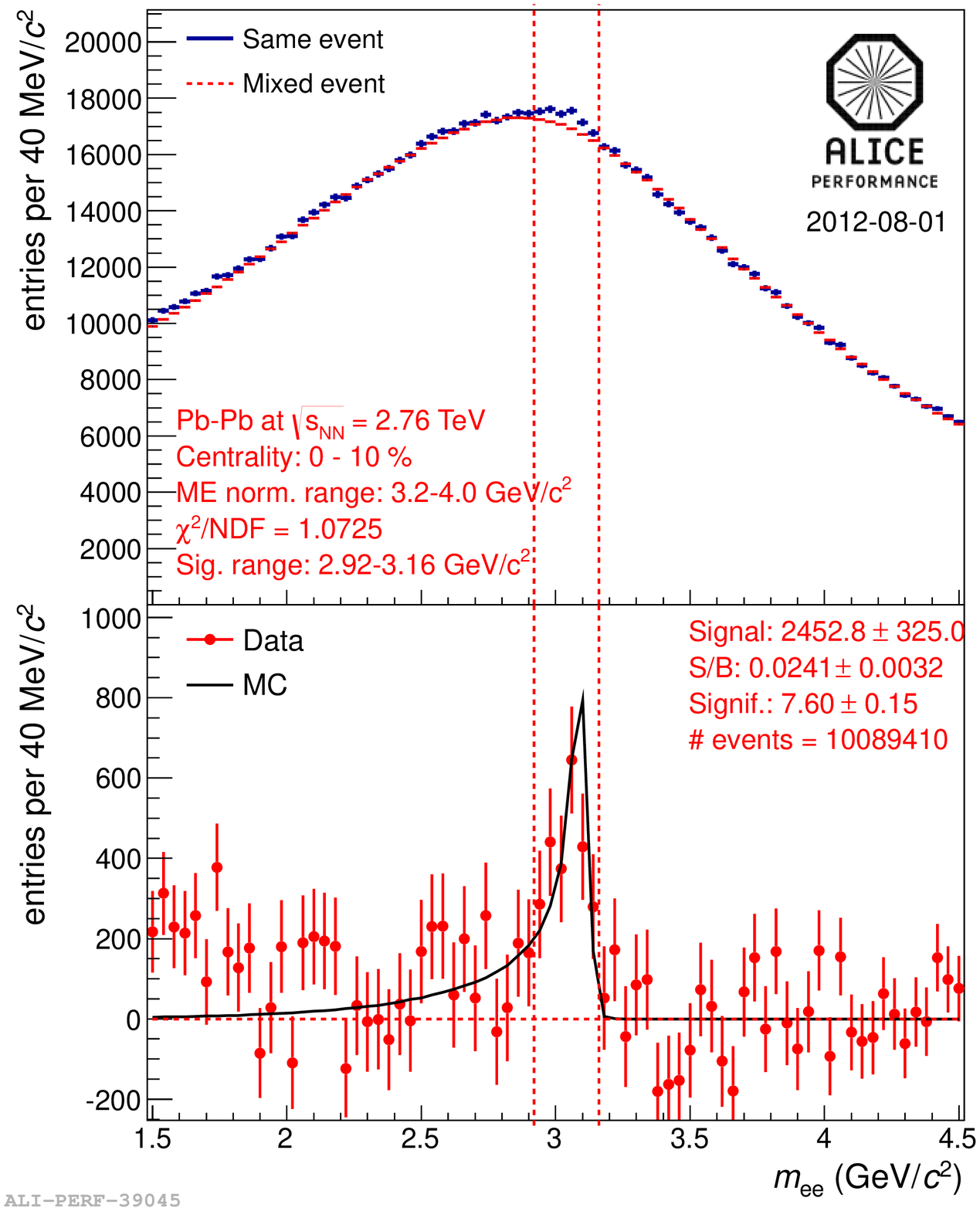}
\end{minipage} \end{tabular}
\caption{\label{J0} Invariant mass distributions measured for the $\mu^+\mu^-$ 
channel (Muon Spectrometer, 2.5$<y<$4, right panel; the dashed line is a fit to 
the background) and e$^+$e$^-$ channel (Central Barrel, $|y|<$0.9, right panel; 
the lower panel is the signal obtained after the mixed-event background subtraction,
with the line the signal form obtained from Monte Carlo simulations).}
\end{figure}

\begin{figure}[htb]
\begin{tabular}{cc} 
\begin{minipage}{.5\textwidth}
\centerline{\includegraphics[width=1.\textwidth]{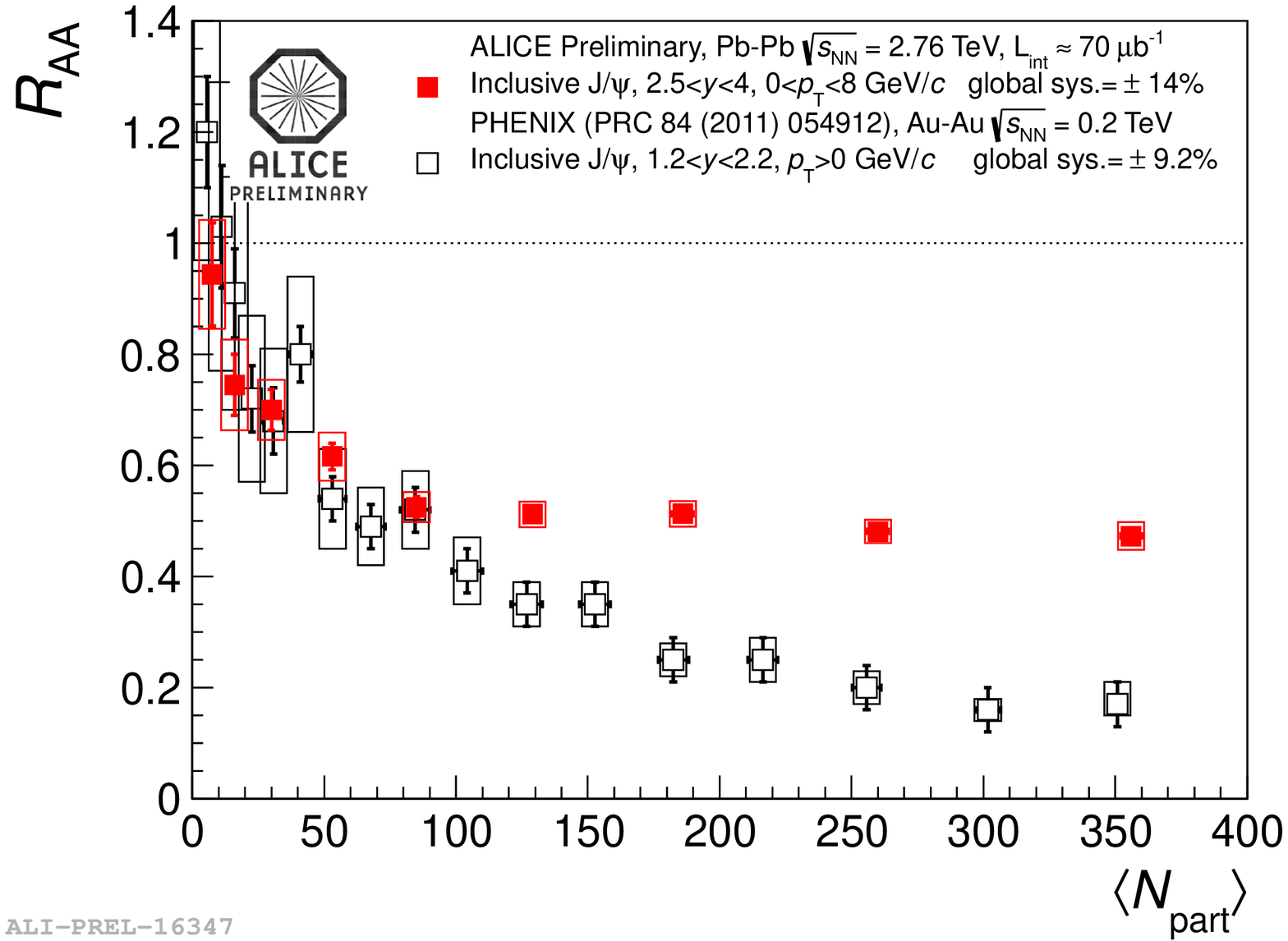}}
\end{minipage} &\begin{minipage}{.5\textwidth}
\hspace{-.6cm}{\includegraphics[width=1.\textwidth]{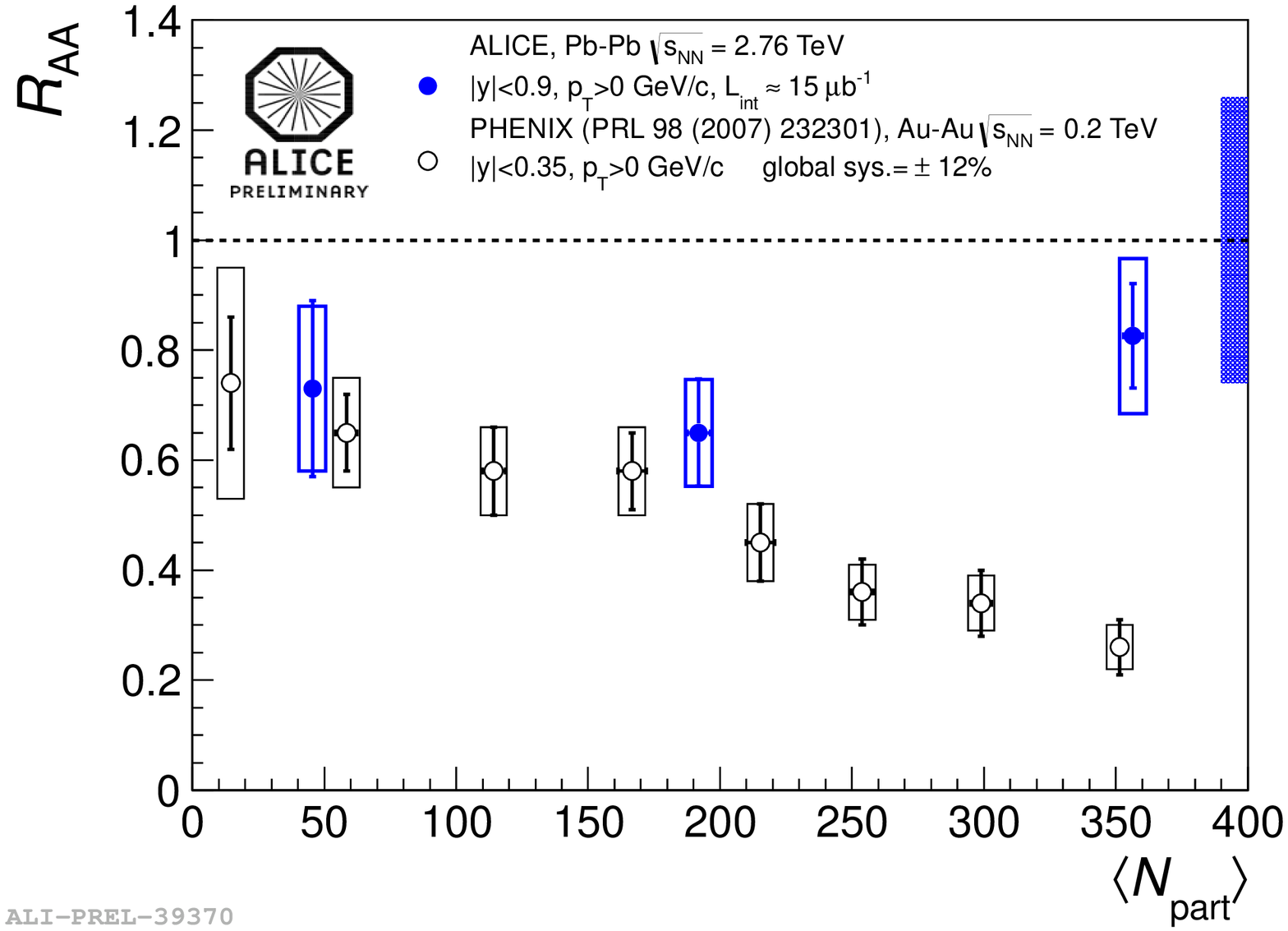}}
\end{minipage} \end{tabular}
\caption{\label{J1} Centrality dependence of the nuclear modification factor 
for inclusive J/$\psi$ production. 
The ALICE measurement (preliminary) at the LHC is compared to the PHENIX 
data at RHIC. The two panels show the data at forward rapidity (left) and
at  mid-rapidity (right).}
\end{figure}

Quarkonium is measured in ALICE via its dielectron channel at midrapidity and
via the dimuon channel at forward rapidity, respectively, as shown
in Fig.~\ref{J0} (see \cite{Abelev:2012rv,Aamodt:2011gj} for the details of 
the analysis).
A clear difference to the RHIC results \cite{Adare:2006ns,Adare:2011yf}
was seen in the first LHC measurement of the overall (inclusive in 
$p_{\mathrm T}$) production, performed in ALICE \cite{Abelev:2012rv}. 
In this case, where low $p_{\mathrm T}$ J/$\psi$ production is dominant, 
less nuclear suppression (larger $R_{\mathrm{AA}}$ value) is seen at the LHC both
at forward rapidity \cite{Suire:2012aa} and at mid-rapidity 
\cite{Arsene:2012uj}, as shown in Fig.~\ref{J1}.
Corroborated with the CMS data \cite{Chatrchyan:2012np}, teh result indicates
an enhanced production of J/$\psi$ at low-$p_{\mathrm T}$ compared to 
high $p_{\mathrm T}$.

At the LHC, the estimated energy density is at least a factor of 3
larger than at RHIC \cite{Chatrchyan:2012mb}, leading to an initial temperature
most likely above the one required for J/$\psi$ dissociation. 
Therefore, one can conclude that the production mechanism of J/$\psi$ 
and charmonium in general at the LHC is determined (to a large extent) 
by regeneration in QGP or by generation at chemical freeze-out. 

\begin{figure}[htb]
\begin{tabular}{cc} 
\begin{minipage}{.5\textwidth}
\centerline{\includegraphics[width=1.\textwidth]{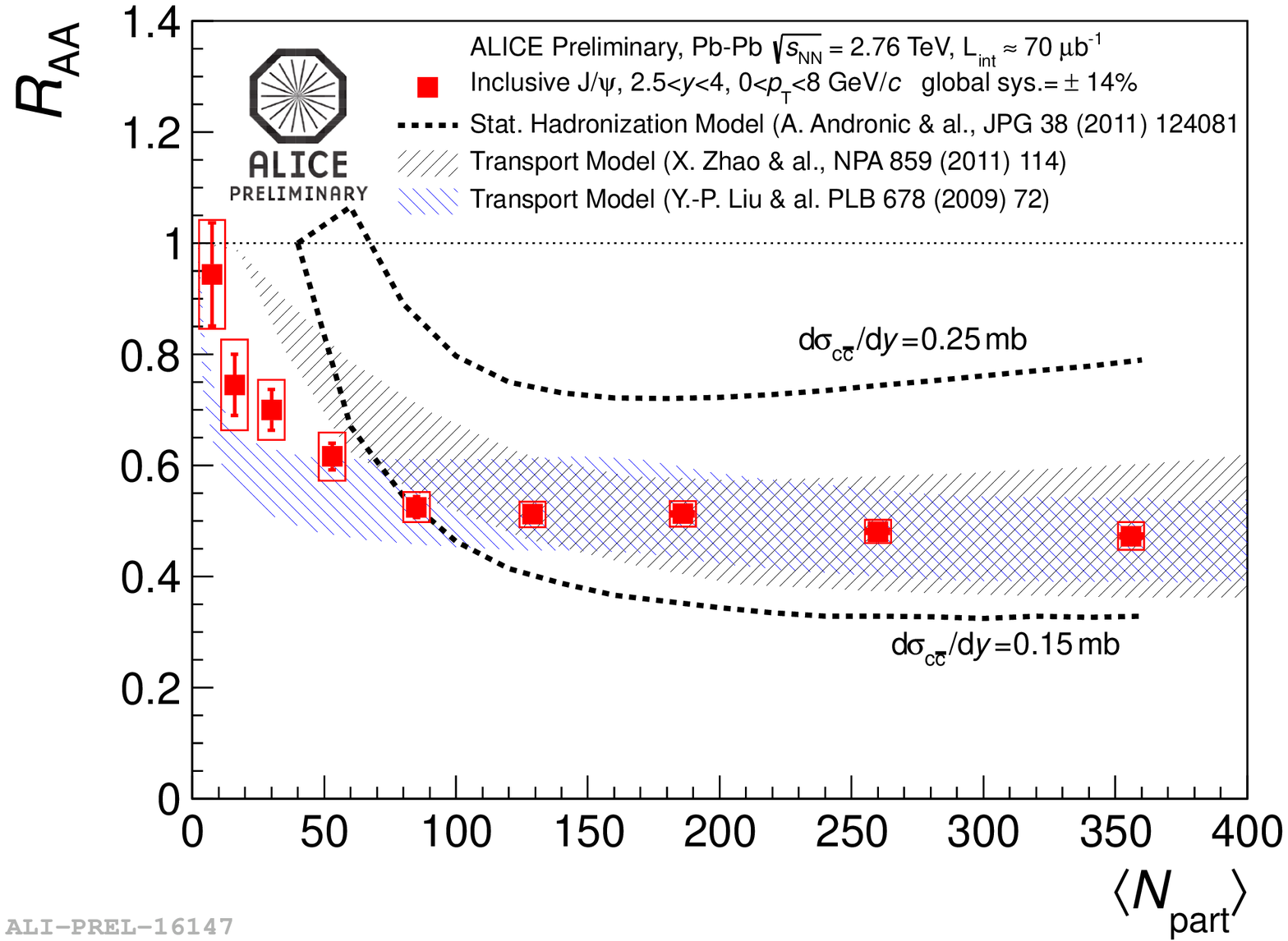}}
\end{minipage} &\begin{minipage}{.5\textwidth}
\hspace{-.6cm}{\includegraphics[width=1.\textwidth]{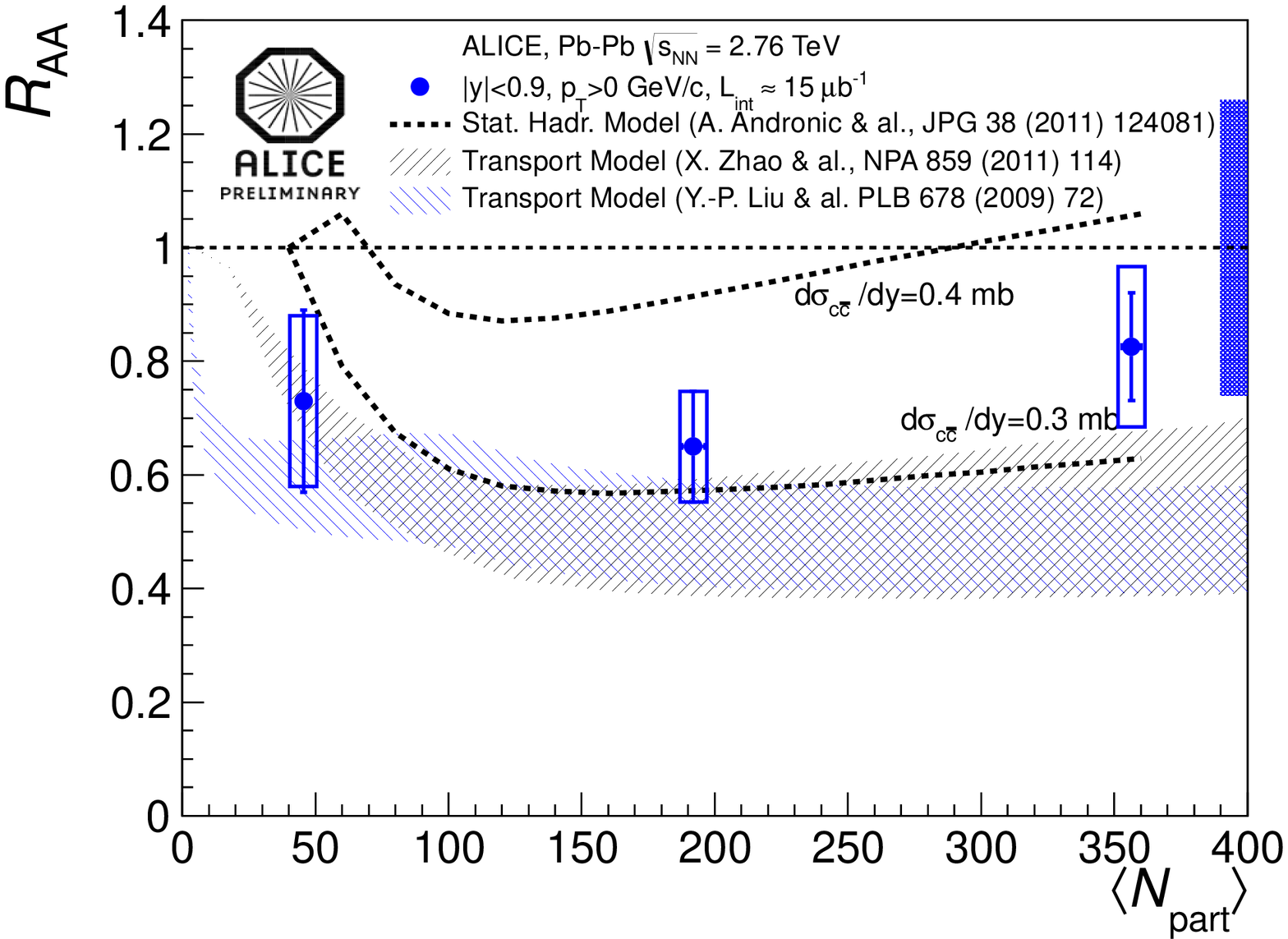}}
\end{minipage} \end{tabular}
\caption{\label{J2} 
Centrality dependence of the nuclear modification factor for 
J/$\psi$ at all momenta in comparison to theoretical models, for forward 
rapidity (left panel) and mid-rapidity (right panel).}
\end{figure}

Indeed, both the statistical hadronization  \cite{Andronic:2011yq} 
and transport  \cite{Liu:2009nb,Zhao:2011cv} models reproduce 
the data \cite{Abelev:2012rv}, as seen in Fig.~\ref{J2}. 
Based on these observations, the J/$\psi$ production can be considered 
a probe of QGP as initially proposed \cite{Matsui:1986dk}, but may not be 
a  ``thermometer'' of the medium \cite{Karsch:2005nk}.
Within the statistical model, the charmonium states become probes 
of the phase boundary between QGP and hadron phase. This extends 
with a heavy quark the family of quarks employed for the determination of 
the hadronization temperature (via the conjectured connection to the 
chemical freeze-out temperature extracted from fits of statistical model 
calculations to hadron abundances).

\begin{figure}[htb]
\centerline{\includegraphics[width=.6\textwidth]{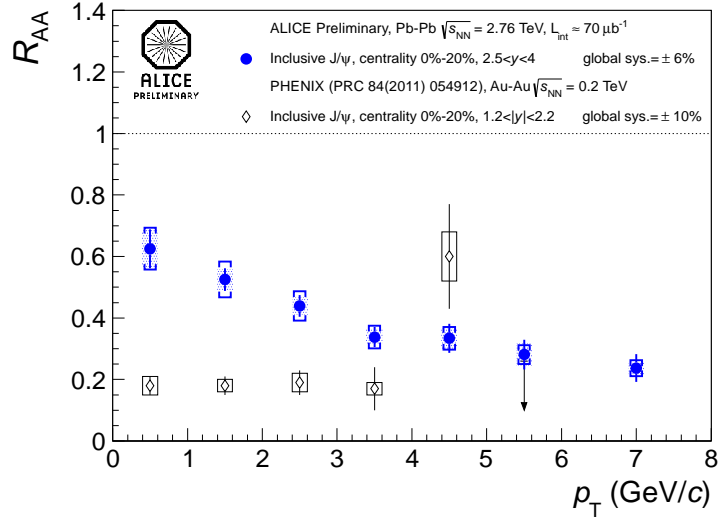}}
\caption{\label{J3}. Transverse momentum dependence of the nuclear modification 
factor for J/$\psi$ at forward rapidity for the centrality range 0-20\%. 
The preliminary ALICE data at the LHC are compared to measurements at RHIC by 
PHENIX.}
\end{figure}

\begin{figure}[htb]
\begin{tabular}{cc} 
\begin{minipage}{.5\textwidth}
\centerline{\includegraphics[width=1.\textwidth]{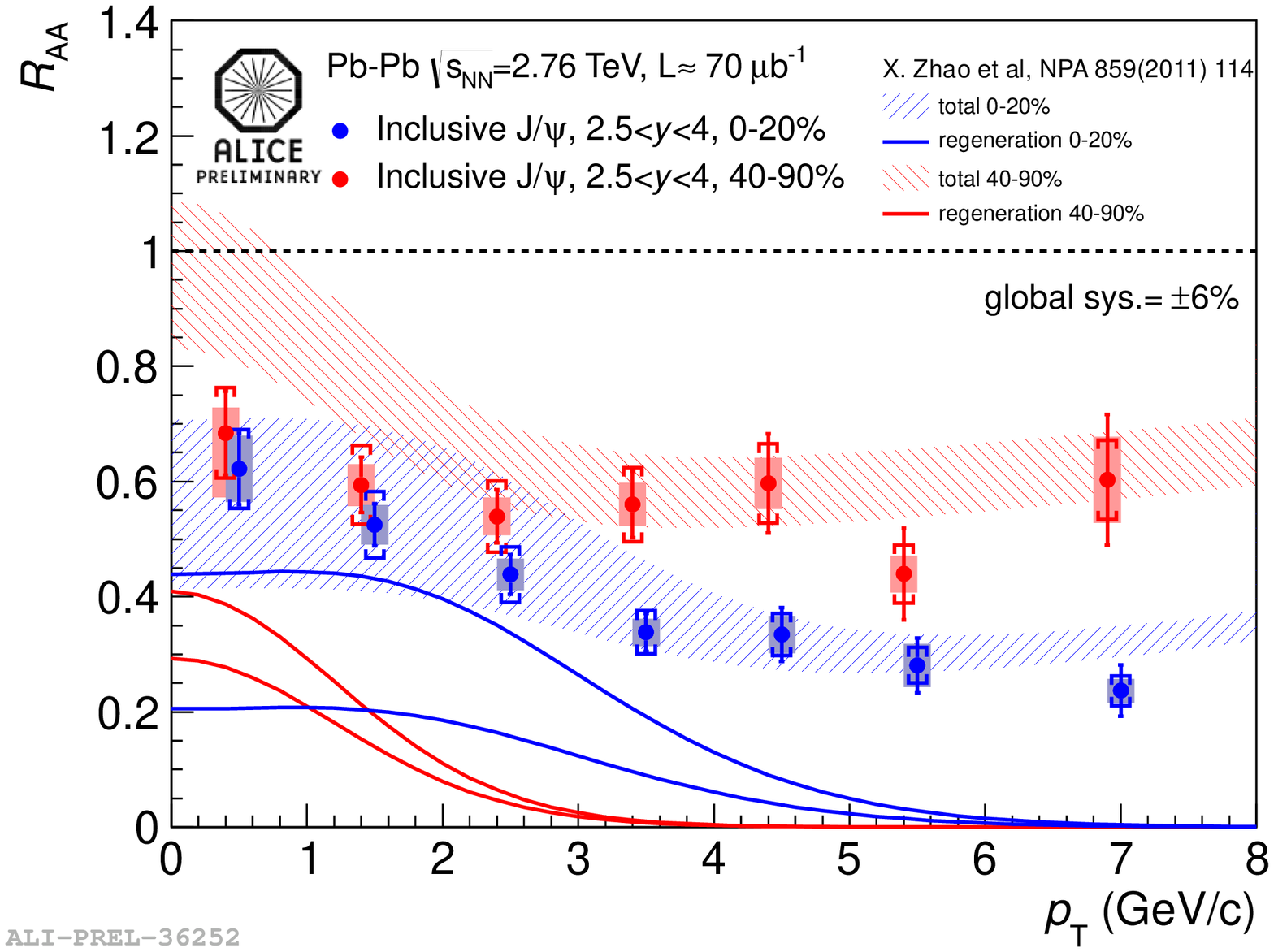}}
\end{minipage} &\begin{minipage}{.5\textwidth}
\hspace{-.6cm}{\includegraphics[width=1.\textwidth]{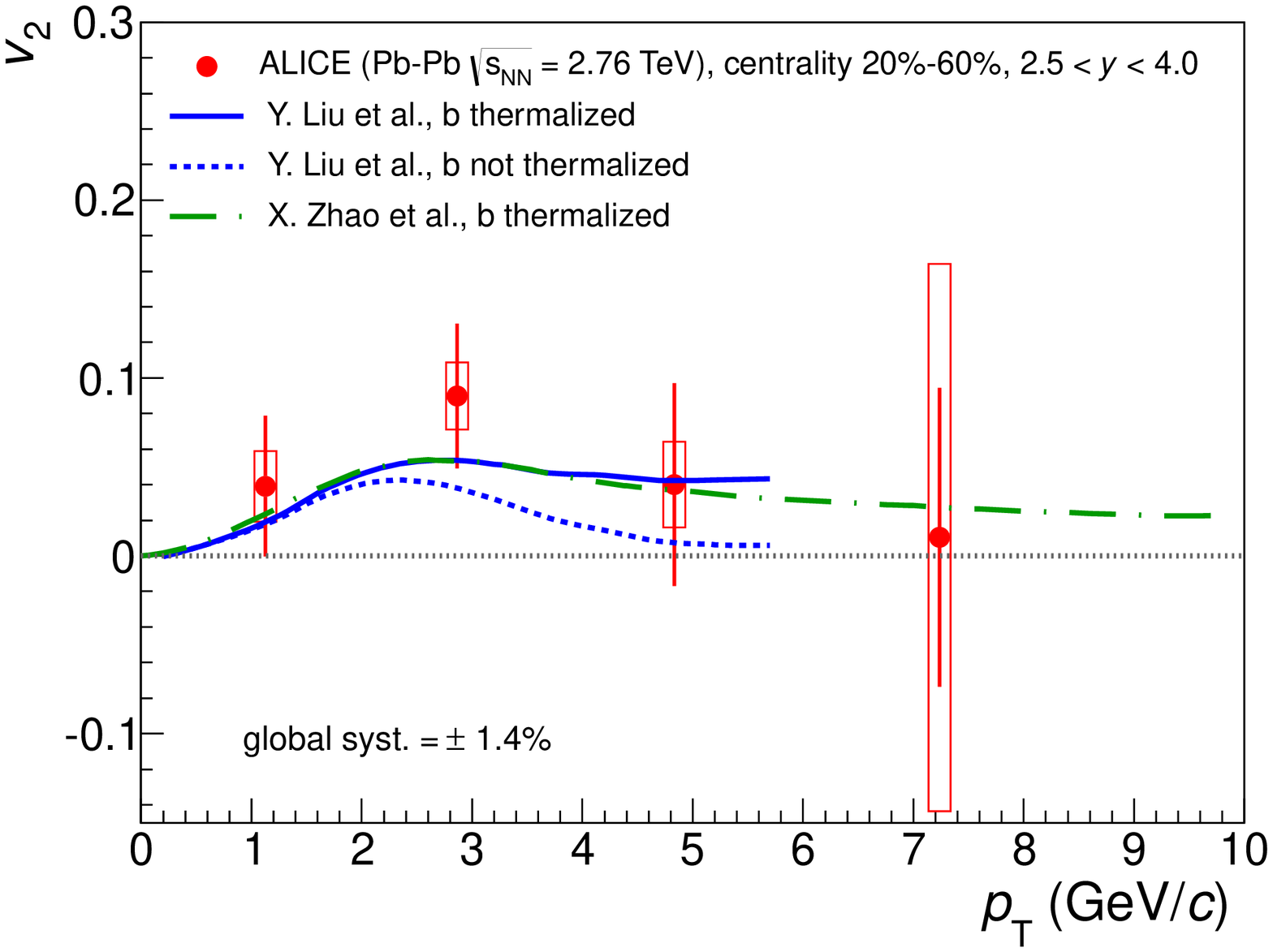}}
\end{minipage} \end{tabular}
\caption{\label{J4} 
Nuclear modification factor (left panel, for centrality 
ranges 0-20\% and  40-90\%) and elliptic flow (right panel, for centrality 
range 20-60\%) of J/$\psi$ mesons as a function of transverse momentum.
The ALICE data is compared with parton transport model predictions.}
\end{figure}

The transverse momentum dependence of the nuclear modification factor,
shown in Fig.~\ref{J3}, is, at the LHC, dramatically different
than the one measured at RHIC. At low-$p_{\mathrm T}$ the nuclear suppression 
is significantly reduced (i.e. larger $R_{\mathrm{AA}}$) at the 
LHC \cite{Suire:2012aa} compared to RHIC \cite{Adare:2006ns}.
Transport model calculations reproduce the data quantitatively, as can be seen 
in Fig.~\ref{J4} (with model of ref. \cite{Zhao:2011cv}; the open bands 
represent the yield due to regeneration).
In current models \cite{Liu:2009nb,Zhao:2011cv}, about half of the 
low-$p_{\mathrm T}$ J/$\psi$ yield in Pb-Pb collisions at 
$\sqrt{s_{\mathrm{NN}}}$ = 2.76 TeV is produced by the recombination of charm 
quarks in QGP, while the rest is due to primordial production. 

Both the kinetic and the statistical hadronization models require 
thermalization of the charm quarks in QGP.
As a consequence they will follow the collective behavior of the bulk
QGP and their flow will be reflected in that of charmed hadrons and quarkonia.
Indeed, elliptic flow of J/$\psi$ at LHC energies was predicted within 
a transport model \cite{Liu:2009nb}.
The first measurement at the LHC \cite{ALICE:2013xna}, 
shown in Fig.~\ref{J4} (right), provides a tantalizing hint of a non-zero
elliptic flow of J/$\psi$ (the significance of non-zero flow for the 
centrality range 20-60\% is 2.3$\sigma$ in the $p_{\mathrm T}$ range 2-4 GeV/$c$). 
The preliminary ALICE data is consistent with the expectation from transport 
models, but precision data is needed in order to be able to extract information
on the QGP properties and on the amount of J/$\psi$ produced via 
regeneration.

The picture outlined above for the charmonium production, extracted by 
comparison of data and model predictions, remains to be tested by precision
measurements at the LHC, which will place further constraints to models. 
A generic feature expected within (re)generation models is an increase
of $R_{\mathrm{AA}}$ at the higher energy ($\sqrt{s_{\mathrm{NN}}}$ = 5.1 TeV) 
of collisions expected in the LHC after the current shutdown period.
In particular, both the statistical and the transport models employ as input 
parameter the $c\bar{c}$ production cross section $\sigma_{c\bar{c}}$.
The sensitivity of the calculations to $\sigma_{c\bar{c}}$ is rather large
(see Fig.~\ref{J2}). 
A precision measurement of $\sigma_{c\bar{c}}$ in Pb-Pb collisions,
within reach with the proposed ALICE upgrade \cite{Musa:1475243}, will place 
an important constraint to models.
A precision measurement of excited states of charmonium, in particular 
of $\psi(2S)$, for which current data \cite{Arnaldi:2012bg} are rather 
imprecise, could allow to discriminate between the regeneration 
\cite{Zhao:2011cv} and statistical production \cite{Andronic:2009sv} mechanisms.

\section{Summary and outlook}
A wealth of data has been measured by ALICE on charmed hadrons and charmonium
in Pb-Pb collisions at the LHC. Parton energy loss is strong also for charm 
quarks and its quantitative description together with the observed elliptic
flow in theoretical models remains a challenge.
Charmonium exhibits features of production via regeneration in the 
hot deconfined medium or via statistical hadronization at the phase boundary.
Disentangling these two pictures is the next challenge for the experiments 
and theory.

\section*{Acknowledgments}
I would like to thank the organizers for the invitation to talk at this special 
workshop at the close proximity to the magnificent Kruger Park and in 
particular many thanks to Simon Connell for a special strategy to obtain an 
unlikely visa at the Johannesburg airport.

\section*{References}
\bibliographystyle{iopart-num}   

\bibliography{Q_Kru}

\providecommand{\newblock}{}
\begin{thebibliography}{10}
\expandafter\ifx\csname url\endcsname\relax
  \def\url#1{{\tt #1}}\fi
\expandafter\ifx\csname urlprefix\endcsname\relax\def\urlprefix{URL }\fi
\providecommand{\eprint}[2][]{\url{#2}}

\bibitem{Collins:1974ky}
Collins J~C and Perry M 1975 {\em Phys. Rev. Lett.\/} {\bf 34} 1353

\bibitem{Cabibbo:1975ig}
Cabibbo N and Parisi G 1975 {\em Phys. Lett. B\/} {\bf 59} 67--69

\bibitem{Karsch:2001cy}
Karsch F 2002 {\em Lect. Notes Phys.\/} {\bf 583} 209--249 (\textit{Preprint}
  \eprint{hep-lat/0106019})

\bibitem{ALICE:2011aa}
Abelev B {\em et~al.\/} (ALICE Collaboration) 2012 {\em JHEP\/} {\bf 1201} 128
  (\textit{Preprint} \eprint{1111.1553})

\bibitem{Kniehl:2012ti}
Kniehl B, Kramer G, Schienbein I and Spiesberger H 2012 {\em Eur. Phys. J. C\/}
  {\bf 72} 2082 (\textit{Preprint} \eprint{1202.0439})

\bibitem{Cacciari:2012ny}
Cacciari M, Frixione S, Houdeau N, Mangano M~L, Nason P {\em et~al.\/} 2012
  {\em JHEP\/} {\bf 1210} 137 (\textit{Preprint} \eprint{1205.6344})

\bibitem{Maciula:2013wg}
Maciula R and Szczurek A 2013  (\textit{Preprint} \eprint{1301.3033})

\bibitem{Miller:2007ri}
Miller M~L, Reygers K, Sanders S~J and Steinberg P 2007 {\em Ann. Rev. Nucl.
  Part. Sci.\/} {\bf 57} 205--243 (\textit{Preprint} \eprint{nucl-ex/0701025})

\bibitem{Heinz:2013th}
Heinz U~W and Snellings R 2013  (\textit{Preprint} \eprint{1301.2826})

\bibitem{Aamodt:2008zz}
Aamodt K {\em et~al.\/} (ALICE Collaboration) 2008 {\em JINST\/} {\bf 3} S08002

\bibitem{Dokshitzer:2001zm}
Dokshitzer Y~L and Kharzeev D 2001 {\em Phys. Lett. B\/} {\bf 519} 199--206
  (\textit{Preprint} \eprint{hep-ph/0106202})

\bibitem{Adare:2010de}
Adare A {\em et~al.\/} (PHENIX Collaboration) 2011 {\em Phys. Rev. C\/} {\bf
  84} 044905 (\textit{Preprint} \eprint{1005.1627})

\bibitem{ALICE:2012ab}
Abelev B {\em et~al.\/} (ALICE Collaboration) 2012 {\em JHEP\/} {\bf 1209} 112
  (\textit{Preprint} \eprint{1203.2160})

\bibitem{delValle:2012qw}
del Valle Z~C (ALICE Collaboration) 2012  (\textit{Preprint}
  \eprint{1212.0385})

\bibitem{Abelev:2012qh}
Abelev B {\em et~al.\/} (ALICE Collaboration) 2012 {\em Phys. Rev. Lett.\/}
  {\bf 109} 112301 (\textit{Preprint} \eprint{1205.6443})

\bibitem{Innocenti:2012ds}
Innocenti G~M (ALICE Collaboration) 2012  (\textit{Preprint}
  \eprint{1210.6388})

\bibitem{He:2012df}
He M, Fries R~J and Rapp R 2013 {\em Phys. Rev. Lett.\/} {\bf 110} 112301
  (\textit{Preprint} \eprint{1204.4442})

\bibitem{Gossiaux:2009mk}
Gossiaux P, Bierkandt R and Aichelin J 2009 {\em Phys. Rev. C\/} {\bf 79}
  044906 (\textit{Preprint} \eprint{0901.0946})

\bibitem{Greco:2011fs}
Greco V, van Hees H and Rapp R 2012 {\em AIP Conf.Proc.\/} {\bf 1422} 117--126
  (\textit{Preprint} \eprint{1110.4138})

\bibitem{Sharma:2009hn}
Sharma R, Vitev I and Zhang B~W 2009 {\em Phys. Rev. C\/} {\bf 80} 054902
  (\textit{Preprint} \eprint{0904.0032})

\bibitem{Alberico:2011zy}
Alberico W, Beraudo A, De~Pace A, Molinari A, Monteno M {\em et~al.\/} 2011
  {\em Eur. Phys. J. C\/} {\bf 71} 1666 (\textit{Preprint} \eprint{1101.6008})

\bibitem{Buzzatti:2011vt}
Buzzatti A and Gyulassy M 2012 {\em Phys. Rev. Lett.\/} {\bf 108} 022301
  (\textit{Preprint} \eprint{1106.3061})

\bibitem{Uphoff:2012it}
Uphoff J, Fochler O, Xu Z and Greiner C 2012  (\textit{Preprint}
  \eprint{1208.1970})

\bibitem{Horowitz:2012cf}
Horowitz W 2012  (\textit{Preprint} \eprint{1210.8330})

\bibitem{Abelev:2012eq}
Abelev B {\em et~al.\/} (ALICE Collaboration) 2013 {\em Phys. Lett. B\/} {\bf
  720} 52 (\textit{Preprint} \eprint{1208.2711})

\bibitem{CMS-PAS-HIN-12-014}
Chatrchyan S {\em et~al.\/} (CMS Collaboration) 2012 {\em Physics Analysis
  Note\/} {\bf CMS-PAS-HIN-12-014}

\bibitem{Banerjee:2011ra}
Banerjee D, Datta S, Gavai R and Majumdar P 2012 {\em Phys. Rev. D\/} {\bf 85}
  014510 (\textit{Preprint} \eprint{1109.5738})

\bibitem{Matsui:1986dk}
Matsui T and Satz H 1986 {\em Phys. Lett. B\/} {\bf 178} 416

\bibitem{Kluberg:2009wc}
Kluberg L and Satz H 2009  (\textit{Preprint} \eprint{0901.3831})

\bibitem{Arnaldi:2010ky}
Arnaldi R {\em et~al.\/} (NA60 Collaboration) 2012 {\em Phys. Lett. B\/} {\bf
  706} 263--267 (\textit{Preprint} \eprint{1004.5523})

\bibitem{Alessandro:2004ap}
Alessandro B {\em et~al.\/} (NA50 Collaboration) 2005 {\em Eur. Phys. J. C\/}
  {\bf 39} 335--345 (\textit{Preprint} \eprint{hep-ex/0412036})

\bibitem{Adare:2006ns}
Adare A {\em et~al.\/} (PHENIX Collaboration) 2007 {\em Phys. Rev. Lett.\/}
  {\bf 98} 232301 (\textit{Preprint} \eprint{nucl-ex/0611020})

\bibitem{Digal:2001ue}
Digal S, Petreczky P and Satz H 2001 {\em Phys. Rev. D\/} {\bf 64} 094015
  (\textit{Preprint} \eprint{hep-ph/0106017})

\bibitem{Karsch:2005nk}
Karsch F, Kharzeev D and Satz H 2006 {\em Phys. Lett. B\/} {\bf 637} 75--80
  (\textit{Preprint} \eprint{hep-ph/0512239})

\bibitem{Alessandro:2006ju}
Alessandro B {\em et~al.\/} (NA50 Collaboration) 2007 {\em Eur. Phys. J. C\/}
  {\bf 49} 559--567 (\textit{Preprint} \eprint{nucl-ex/0612013})

\bibitem{Sorge:1997bg}
Sorge H, Shuryak E~V and Zahed I 1997 {\em Phys. Rev. Lett.\/} {\bf 79}
  2775--2778 (\textit{Preprint} \eprint{hep-ph/9705329})

\bibitem{BraunMunzinger:2000px}
Braun-Munzinger P and Stachel J 2000 {\em Phys. Lett. B\/} {\bf 490} 196--202
  (\textit{Preprint} \eprint{nucl-th/0007059})

\bibitem{Andronic:2006ky}
Andronic A, Braun-Munzinger P, Redlich K and Stachel J 2007 {\em Nucl. Phys.
  A\/} {\bf 789} 334--356 (\textit{Preprint} \eprint{nucl-th/0611023})

\bibitem{Andronic:2007bi}
Andronic A, Braun-Munzinger P, Redlich K and Stachel J 2007 {\em Phys. Lett.
  B\/} {\bf 652} 259--261 (\textit{Preprint} \eprint{nucl-th/0701079})

\bibitem{BraunMunzinger:2009ih}
Braun-Munzinger P and Stachel J 2009  (\textit{Preprint} \eprint{0901.2500})

\bibitem{Adare:2011yf}
Adare A {\em et~al.\/} (PHENIX Collaboration) 2011 {\em Phys. Rev. C\/} {\bf
  84} 054912 (\textit{Preprint} \eprint{1103.6269})

\bibitem{Andronic:2011yq}
Andronic A, Braun-Munzinger P, Redlich K and Stachel J 2011 {\em J. Phys. G\/}
  {\bf 38} 124081 (\textit{Preprint} \eprint{1106.6321})

\bibitem{Andronic:2009sv}
Andronic A, Beutler F, Braun-Munzinger P, Redlich K and Stachel J 2009 {\em
  Phys. Lett. B\/} {\bf 678} 350--354 (\textit{Preprint} \eprint{0904.1368})

\bibitem{Thews:2000rj}
Thews R~L, Schroedter M and Rafelski J 2001 {\em Phys. Rev. C\/} {\bf 63}
  054905 (\textit{Preprint} \eprint{hep-ph/0007323})

\bibitem{Liu:2009nb}
Liu Y~P, Qu Z, Xu N and Zhuang P~F 2009 {\em Phys. Lett. B\/} {\bf 678} 72--76
  (\textit{Preprint} \eprint{0901.2757})

\bibitem{Zhao:2011cv}
Zhao X and Rapp R 2011 {\em Nucl. Phys. A\/} {\bf 859} 114--125
  (\textit{Preprint} \eprint{1102.2194})

\bibitem{Aad:2010aa}
Aad G {\em et~al.\/} (ATLAS Collaboration) 2011 {\em Phys. Lett. B\/} {\bf 697}
  294--312 (\textit{Preprint} \eprint{1012.5419})

\bibitem{Chatrchyan:2012np}
Chatrchyan S {\em et~al.\/} (CMS Collaboration) 2012 {\em JHEP\/} {\bf 1205}
  063 (\textit{Preprint} \eprint{1201.5069})

\bibitem{Abelev:2012rv}
Abelev B {\em et~al.\/} (ALICE Collaboration) 2012 {\em Phys. Rev. Lett.\/}
  {\bf 109} 072301 (\textit{Preprint} \eprint{1202.1383})

\bibitem{Aamodt:2011gj}
Aamodt K {\em et~al.\/} (ALICE Collaboration) 2011 {\em Phys. Lett. B\/} {\bf
  704} 442--455 (\textit{Preprint} \eprint{1105.0380})

\bibitem{Suire:2012aa}
Suire C (ALICE collaboration) 2012 {\em Hard Probes\/} (\textit{Preprint}
  \eprint{1208.5601})

\bibitem{Arsene:2012uj}
Arsene I~C (ALICE Collaboration) 2012 {\em Nucl.Phys.A\/} (\textit{Preprint}
  \eprint{1210.5818})

\bibitem{Chatrchyan:2012mb}
Chatrchyan S {\em et~al.\/} (CMS Collaboration) 2012 {\em Phys. Rev. Lett.\/}
  {\bf 109} 152303 (\textit{Preprint} \eprint{1205.2488})

\bibitem{ALICE:2013xna}
Abbas E {\em et~al.\/} (ALICE Collaboration) 2013  (\textit{Preprint}
  \eprint{1303.5880})

\bibitem{Musa:1475243}
Musa L and Safarik K 2012 Letter of intent for the upgrade of the alice
  experiment Tech. Rep. CERN-LHCC-2012-012. LHCC-I-022 CERN Geneva

\bibitem{Arnaldi:2012bg}
Arnaldi R (ALICE Collaboration) 2012  (\textit{Preprint} \eprint{1211.2578})

\end{thebibliography}

\IfFileExists{\jobname.bbl}{}
 {\typeout{}
  \typeout{******************************************}
  \typeout{** Please run "bibtex \jobname" to optain}
  \typeout{** the bibliography and then re-run LaTeX}
  \typeout{** twice to fix the references!}
  \typeout{******************************************}
  \typeout{}
 }

\end{document}